# On the mathematical description of polarization in optical communications and how their induced impairments can be minimized

## CARLOS L. JANER

*Departamento de Ingeniería Electrónica, Escuela Superior de Ingenieros, Sevilla 41092, Spain*
*janer@us.es*

**In this paper, it is claimed that the correct mathematical framework of combined polarization mode dispersion and polarization dependent losses (combined PMD-PDL effects or impairments) in optical fibers is the irreducible spinor representation of the Lorentz Group. Combined PMD-PDL effects are shown to be formally identical to Lorentz Transformations acting on spin ½ zero mass particles. Since there are two different irreducible spinor representations of the restricted Lorentz Group, there must also exist two kinds of states of polarization (SOPs) that are relevant in the description of PMD-PDL effects. The optical process that converts one kind into the other is identified as optical phase conjugation. Optical phase conjugation plays the same role as the time inversion operator in the Lorentz Group representation theory. A practical and important example of utility of these ideas, a technique that significantly reduces the PMD-PDL induced impairments, is presented. The proposed mathematical model is explained in physical terms and a simple experimental setup designed to either validate or falsify the model is proposed.**

## 1. INTRODUCTION

Polarization plays a very important role in optical communications. Single mode fibers support two orthogonally polarized modes that are fully degenerate. In Polarization Multiplexed (PM) optical communication systems [1], these two modes are used to transmit information coming from two independent sources. The optical carrier is usually modulated both in amplitude and phase to increase the capacity of the communication channel. As light propagates along the fiber, random birefringence breaks the symmetry of the two degenerate modes, induces an energy exchange between these two modes and produces, in sufficiently long fibers, polarization mode dispersion (PMD). PMD introduces crosstalk and a differential delay between the two modes that carry the two multiplexed signals. In coherent detection schemes, PMD induced impairments can be compensated with digital processing (DSP) techniques [2].

Polarization dependent loss (PDL) typically arises in optical components such as optical isolators and introduces in optical systems (where attenuation is usually compensated with Erbium Doped Fiber Amplifiers, EDFAs) polarization dependent loss and gain. PDL cannot be compensated in coherent receivers by DSP processing and is the most important source of system degradation [3]. PDL induces polarization dependent power fluctuations resulting in unequal optical signal-to-noise ratio (OSNR) on each received polarization.

When the fiber PMD is intertwined with PDL elements, the resulting polarization effects are more complicated than PDL or PMD alone and the communication impairments can be, and usually are, more severe than either isolated effect. An accurate mathematical description of these effects is crucial in Polarization Multiplexed coherent optical communication systems.

The main purpose of this paper is to show that the mathematical formulation that correctly describes polarization issues in optical communication is the spinor irreducible representation theory of the Lorentz Group. Although the idea of the Lorentz Group describing polarization transformations in Optics is well known, [4]-[6], it has never, to the best of my knowledge, been considered relevant to the field of optical communications. In this paper it will be analyzed why this happened, it will be proven that not taking into account these ideas was a serious mistake and it will be shown that polarization dependent (combined PMD-PDL) impairments can be easily reduced (the PDL part of the impairments can be minimized) once the full implications of this mathematical formulation are understood. The main issue that has been ignored in this field is that there are two different kinds of states of polarization (SOPs) (that are mutually related by optical phase conjugation) that correspond to the two irreducible spinor representations of the Lorentz Group. These two kinds of SOPs carry all the relevant information about how combined PMD-PDL affects the polarization states of the propagated waves. Optical phase conjugation is identified as the optical process implementing the time inversion symmetry of the Lorentz Group and connecting the two different kinds of SOPs that must be considered in optical communications. The PDL part of the impairments is minimized when both kinds of SOPs are considered because, as it will be proven in this paper, the conjugated SOPs remain orthogonal at both ends of the fiber and, therefore, must be related by a pure rotation.

## 2. PMD AND PDL IN OPTICAL COMMUNICATIONS

In this section, for the sake of clarity and completeness, a very sketchy introduction of polarization dependent penalties in Optical Communications will be made. This section introduces the mathematical notation that will be used in subsequent sections. Due to the complexity of this subject, a thorough introduction to these effects cannot be made here. For a more detailed introduction to polarization mode dispersion (PMD) and polarization dependent losses (PDL) effects in optical fibers the reader is referred to [7] and specially to chapter 8 of the monograph book [8] from which most of the information of this section has been taken.

The reader should bear in mind that the state of polarization (SOP) of the transmitted wave depends on the complex valued information of the two channels that are being multiplexed. These two channels correspond to the two entries of the spinor (that is to say, the two components of the SOP vector). In PM coherent systems the transmitted information is encoded in the instantaneous SOP of the wave. This state of polarization will be in the rest of the paper indistinctly designated as "spinor" or SOP.

There always exists an orthogonal pair of polarization states at the output of a lossless concatenation of birefringent elements which are stationary to first order in frequency. These two states are called the Principal States of Polarization (PSPs). A differential delay exists between signals launched along one PSP (slow PSP) and its orthogonal complement (fast PSP). The pointing direction of the PMD vector is aligned to the slow PSP, the PSP that imparts more delay than the other. The length of the PMD vector is the differential-group delay between the slow and fast PSPs. The traceless Hermitian operator (or matrix) whose eigenvectors define the PSPs is [7]:

$$\mathrm{H_r} = \mathrm{j}\mathrm{U}_\omega \mathrm{U}^\dagger = \frac{1}{2}\,(\vec{\tau}.\vec{\sigma}) \tag{1}$$

where U is the 2x2 unitary matrix that defines the birefringence of the fiber. The equation that describes how the output SOPs change with frequency is [7]:

$$|\mathrm{t}\rangle_\omega = \exp[-\tfrac{1}{2}\mathrm{j}\,(\vec{\tau}.\vec{\sigma})\omega]|\mathrm{t}\rangle_0 \tag{2}$$

The eigenvalue-eigenvector equation of (1) states that the output polarization rotates about the principal-state of the fiber even if very slight changes in the optical frequency are considered. Only if the output state is aligned alongside one of the PSPs its state will not change with frequency. The rotation angle depends on the value of the frequency excursion as: $\vec{\varphi} = \omega\,\vec{\tau}$ and the Unitary SU(2) rotation matrix can be obtained from the traceless 2x2 Hermitian matrix (1) by the exponential map [9]-[10] as expressed in (2). From this point of view, the PMD vector generates rotations of the output polarization state and, if we choose to use Jones vectors to describe polarization states, we can regard these transformations as rotations of spin ½ particles [7].

The resulting polarization effects of combined PMD and PDL are more complicated. The principal states of polarization still exist but they are not orthogonal to one another and the output polarization state does not follow a simple rotational motion around the PSPs as a function of frequency [8]. The PSPs are defined as the eigenstates of the following operator (matrix), that is traceless, but not Hermitian ($\mathrm{H_r}$ is Hermitian and $\mathrm{jH_i}$ is skew-Hermitian) [8]:

$$\mathrm{H_r} + \mathrm{jH_i} = \mathrm{j}\mathrm{T}_\omega \mathrm{T}^{-1} = \frac{1}{2}\left(\overrightarrow{\Omega_\mathrm{r}}.\vec{\sigma} + \mathrm{j}\overrightarrow{\Omega_\mathrm{i}}.\vec{\sigma}\right) \tag{3}$$

The vectors $\overrightarrow{\Omega_\mathrm{r}}$ and $\overrightarrow{\Omega_\mathrm{i}}$ are real Stokes vectors (whose components are the six free parameters that characterize the polarization properties of the fiber) that relate to the system birefringence and differential attenuation, but not in a straightforward way [8]. In this analysis it is assumed that there are no isotropic effects, no common gain or loss or, at least, the optical amplifiers exactly compensate for the SOP independent loss. A general 2x2 complex valued matrix whose determinant is equal to 1 can always be generated by a couple of Hermitian and skew-Hermitian given by (3). The exponential of (3) generates the desired matrix.

The complex eigenvalues of this operator are of equal module and opposite signs as a direct consequence of it having a zero trace. If we denote the eigenvalues as: $\lambda = \pm(\tau + \mathrm{j}\eta)$, then the real part $\tau$ is the familiar differential-group delay magnitude; the imaginary part $\eta$ is the differential-attenuation slope (DAS), which is the frequency derivative of the differential attenuation along the two eigenvectors [8]. In the presence of PDL, the PSPs are not orthogonal due to the complex value of the eigenvectors [8]. The equation that describes how the output SOPs change with frequency is:

$$|\mathrm{t}\rangle_\omega = \exp[-\tfrac{1}{2}\mathrm{j}\left(\overrightarrow{\Omega_\mathrm{r}}.\vec{\sigma} + \mathrm{j}\overrightarrow{\Omega_\mathrm{i}}.\vec{\sigma}\right)\omega]|\mathrm{t}\rangle_0 \tag{4}$$

This equation describes the combined rotation and Lorentz boost (a proper Lorentz transformation) of a zero-mass spin ½ particle [9]-[12]. So, the evolution of the SOPs at the fiber output is formally the Lorentz transformation of a two-component spinor. Only if the output state is aligned alongside one of the PSPs its state will not change with frequency. The Lorentz angle depends on the value of excursion frequency as: $\vec{\varphi} = \frac{1}{2}\left[\left(\overrightarrow{\omega\Omega_\mathrm{r}} + \mathrm{j}\,\overrightarrow{\omega\Omega_\mathrm{i}}\right).\vec{\sigma}\right]$ and the Lorentz transformation can be obtained from the traceless 2x2 matrix (3) by the exponential map as shown in (4). From this point of view, the combined PMD and PDL vector generates a restricted Lorentz transformation of the output polarization states and, if we choose to use Jones vectors to describe polarization states, we can regard these as a proper (restricted) Lorentz transformation of spin ½ particles [9]-[12].

The reader should bear in mind that the previous observation is almost trivial. The polarization properties of an optical fiber are described by a very general group of matrices, those 2x2 complex valued matrices whose determinants are equal to 1 that are known in Group Theory [9]-[10] as the rank 2 complex special linear group of matrices, SL(2,C). Since there is a close relationship between the proper Lorentz group, $\mathrm{SO}^+(1,3)$, and SL(2,C) (SL(2,C) is the double cover of $\mathrm{SO}^+(1,3)$) this seems to be just a mathematical curiosity devoid of any practical interest. Possibly this is the reason why, although this relationship is known in Optics [4]-[6], it has never attracted any attention in the field of optical communications. In the following sections it will be show that this was a mistake because combined PMD-PDL induced impairments can be easily alleviated (the PDL part can be minimized) when this fact and its precise mathematical formulation are considered. A crucial part of the mathematical formulation will be to enlarge or extend the restricted Lorentz Group to the general Lorentz Group incorporating the discrete symmetries and identifying the optical processes that implement them. This is, to the best of my knowledge, a novel contribution of this paper.

## 3. IRREDUCIBLE REPRESENTATIONS OF THE LORENTZ GROUP

Equations (3) and (4) show that the six free parameters that define the combined PMD and PDL properties of a fiber are formally equivalent to the six parameters that define a restricted Lorentz transformation of a spin ½ particle [11]-[12]. The Representation Theory of the Lorentz Group is of common knowledge in the field of Particle Physics [9]-[12]. This theory shows that the restricted Lorentz Group has two representations of dimension 2, IRs, which are not equivalent that are denoted as $\mathbf{D}^{(1/2,0)}$ and $\mathbf{D}^{(0,1/2)}$. This can be traced back to the fact that a matrix belonging to the SL(2,C) group (the rank two complex group of matrices with determinant equal to 1), and its complex conjugate, A and $A^*$ are not equivalent in the full group (they are equivalent only in the special unitary, SU(2), subgroup) because one cannot find a similarity transformation relating them [9]-[12]. These matrices constitute two nonequivalent irreducible representations of the restricted Lorentz Group and from these two any other higher order representation (either spinor or tensor) can be built. The spinor IRs are more fundamental than the tensor IRs. From the mathematical point of view, it makes little sense to use Stokes space, $\mathbf{D}^{(1/2,1/2)} = \mathbf{D}^{(1/2,0)} \oplus \mathbf{D}^{(0,1/2)}$ in representation theory, to describe combined PMD and PDL issues in optical communications without any explicit mention to the two IRs, $\mathbf{D}^{(1/2,0)}$ and $\mathbf{D}^{(0,1/2)}$, upon which it is built.

We have then two nonequivalent bases, and, in general, two kinds of spinors (spin ½ particles) denoted by ξ and $\xi^*$ that transform according to following transformation matrices:

$$\mathbf{A} = \mathbf{exp}\left[-\frac{1}{2}\mathbf{j}\left(\overrightarrow{\mathbf{\Omega}_r}.\vec{\boldsymbol{\sigma}} + \mathbf{j}\overrightarrow{\mathbf{\Omega}_I}.\vec{\boldsymbol{\sigma}}\right)\right] \qquad (5)$$

and the representation provided by $A^*$ is equivalent to:

$$\mathbf{B} = \mathbf{exp}\left[-\frac{1}{2}\mathbf{j}\left(\overrightarrow{\mathbf{\Omega}_r}.\vec{\boldsymbol{\sigma}} - \mathbf{j}\overrightarrow{\mathbf{\Omega}_I}.\vec{\boldsymbol{\sigma}}\right)\right] \qquad (6)$$

The irreducible representations of the extended Lorentz Group can be obtained from the restricted Lorentz Group by inclusion of the discrete symmetries $I_S$ (parity inversion), $I_T$ (temporal inversion) and $I_C = I_{ST}$ (inversion symmetry) [12]. These discrete symmetries are very important because they transform one kind of spinor into the other (they connect the two representations) and the optical processes implementing them should be identified. This is a novel contribution of this paper since, to the best of my knowledge, the discrete symmetries of the extended Lorentz Group had never been considered in Optics before and neither had the optical processes implementing them been identified. The fact that these discrete symmetries correspond to optical processes is what makes this mathematical formulation useful and the fact that they had not been identified before as such helps to explain why this mathematical formulation, although known in Optics, had not attracted any interest in the field of optical communications.

### 3.1. TIME INVERSION

The time inversion symmetry affects the spinors in the following way [12]: $I_T$: ξ → $\eta^\dagger$ and $I_T$: $\eta^\dagger$ → -ξ. It transforms the regular contravariant spinor ξ into the conjugate covariant spinor $\eta^\dagger$. If a contravariant spinor ξ transforms as Aξ then its covariant spinor η transforms as $\eta(A^{-1})^T$ (this follows the standard notation in which covariant spinors are represented by row vectors acting on the left of matrices) so time inversion conjugates the spinor and changes "variance" (from either contravariant to covariant or from covariant to contravariant). In optical links that use EDFAs to compensate for the fiber attenuation, covariant spinors would not be allowed to propagate along the whole fiber due to the presence of the isolators placed at both ends of the optical amplifiers (these isolators are placed to prevent these amplifiers from lasing).

The covariant spinor η associated to ξ is given by the expression [12]: $\eta^T = C\xi$ (or $\eta = \xi^T C$) where C is the antisymmetric matrix:

$$C = \begin{pmatrix} 0 & 1 \\ -1 & 0 \end{pmatrix}$$

The optical process that implements this discrete symmetry is closely related to optical phase conjugation [13]-[14]. This optical process would reverse, in counter-propagation, the polarization distortion introduced by the combined effect of PMD and PDL if it were not for the presence of isolators and circulators**.** The device that performs this transformation, that from now on will be called optical conjugation, is a π/4 Faraday rotator followed by an ideal vector phase conjugating mirror [15] and it will be referred to as an optical conjugator. An optical conjugator, therefore, implements the polarization transformation:

$$\xi = \begin{pmatrix} \xi_1 \\ \xi_2 \end{pmatrix} \rightarrow \eta^\dagger = \begin{pmatrix} \xi_2^* \\ -\xi_1^* \end{pmatrix} \qquad (7)$$

Please, notice that the original and transformed SOPs are mutually orthogonal, that is**,** $\langle \eta | \xi \rangle = 0$ all along the fiber.

### 3.2. FRAME INVERSION

The frame inversion transforms spinors in the following way [12]: $I_{ST}$: ξ → jξ and $I_{ST}$: $\eta^\dagger$ → -j $\eta^\dagger$ would not change the SOP and, therefore, would be very difficult to detect. Although it is tempting to speculate and try to assign physical relevance to the "j" factor, it would probably be unjustified. The frame inversion operator (or charge conjugation operator) will be considered as the identity operator in this paper.

## 3.3. PARITY INVERSION

The parity inversion transforms the spinors as follows: $I_S$: $\xi \rightarrow j\eta^{\dagger}$ and $I_S$: $\eta^{\dagger} \rightarrow j\xi$ [12]. Since the frame inversion operator is the identity, the optical device that implements parity inversion is an optical conjugator.

## 4. APPLICATION: PDL CANCELATION

The purpose of this paper is to show that the natural mathematical framework that describes polarization issues in optical communications is the irreducible spinor representation of the Lorentz Group. Since it is a mathematically complex matter, the only convincing way to prove its need is to either solve some complicated problem using it or propose something new and useful based on it. It is now straightforward to do so.

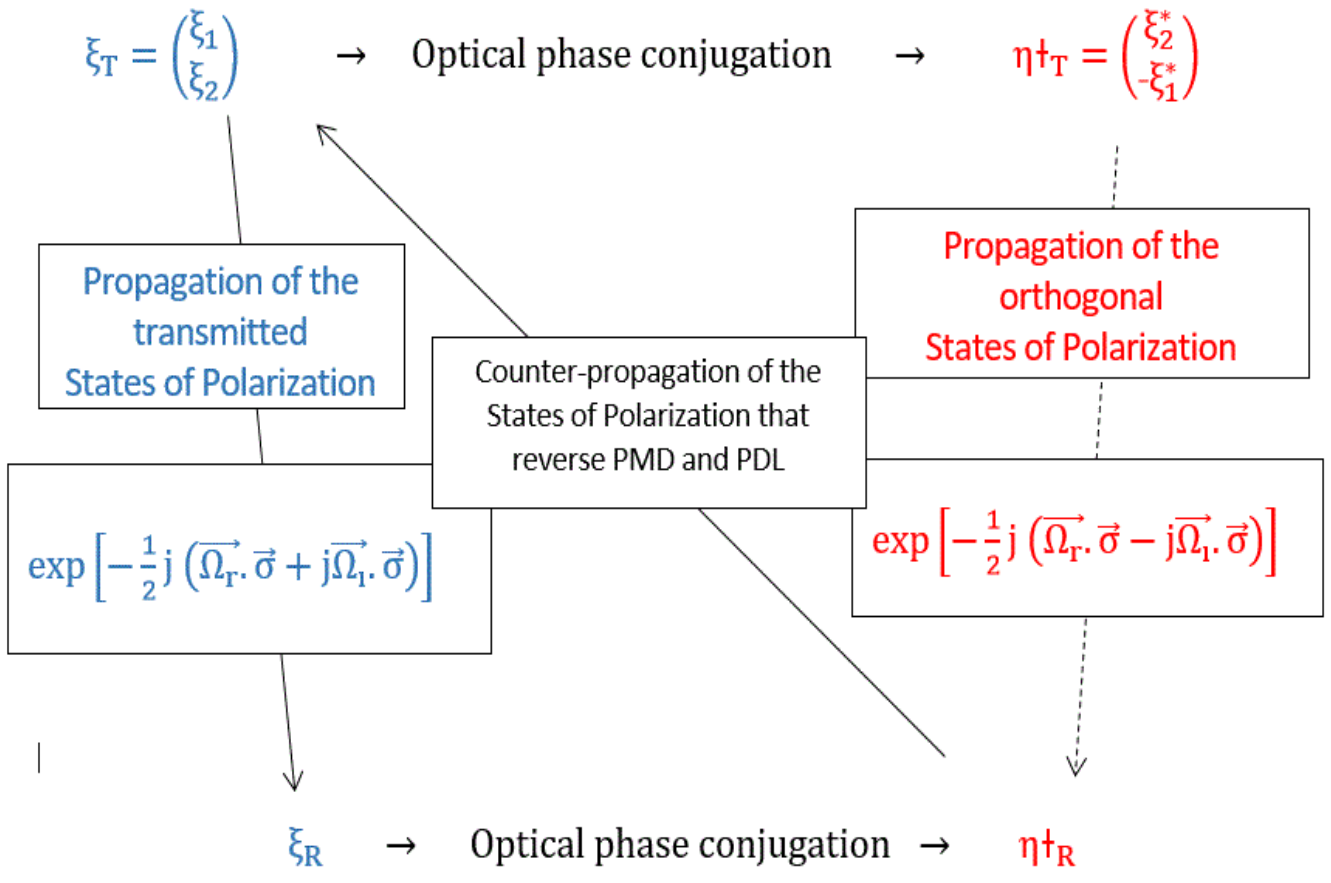

**Figure 1**: Lorentz transformations of the transmitted SOP (or spinor) and its orthogonal SOP.

At the receiver end of the optical link, the received spinor, the SOP of the received wave, $\boldsymbol{\xi_R}$ , is used to estimate what the transmitted spinor was, $\boldsymbol{\xi_T}$. The combined PMD and PDL effects and noise make this estimation process difficult. The previous sections make it clear that we can perform a time inversion transformation at the receiver end of the fiber by means of an optical phase conjugator to obtain $\boldsymbol{\eta_R}^{\dagger}$. If this spinor could counter-propagate along the fiber, it would end at the fiber input as transmitted SOP $\boldsymbol{\xi_T}$. However, as is shown in Figure 1, this spinor can also be considered as the SOP of a propagated wave. Our mathematical framework allows us to identify it with the unique input SOP perpendicular to $\boldsymbol{\xi_T}$. Please, notice that perpendicular spinors belong to conjugated representations and therefore transform differently.

It should be noticed that the sequence of symbols transmitted through the two-mode fiber (the two components of the spinor $\boldsymbol{\xi_T}$ ) is completely random since it corresponds to the information carried by the two multiplexed channels, and therefore, the orthogonal SOPs can only be determined after they have been generated by the transmitter. The propagation depicted in red in Figure 1 does not actually happen in the optical fiber. It is the physical interpretation of considering $\boldsymbol{\eta_R}^{\dagger}$ the SOP of a propagated wave.

**$\boldsymbol{\eta_R}^{\dagger}$ ARE THE SPINORS THAT WOULD BE MEASURED AT THE RECEIVER END, IF THE "ORTHOGONAL TO $\boldsymbol{\xi_T}$ " SPINOR HAD BEEN TRANSMITTED INSTEAD OF $\boldsymbol{\xi_T}$.**

Both spinors, $\xi_R$ and $\eta_R{}^{\dagger}$ can and should be used to estimate the transmitted information and that is exactly what this paper proposes. In this restricted sense, it is as if we had received two different spinors that can be used to estimate the information of a single transmitted symbol, significantly improving the estimation process. Notice also that $\xi$ and $\eta^{\dagger}$ remain orthogonal at both ends of the fiber optic, so that when the joint motion of both spinors is considered**,** the PDL penalty part is minimized**.** This stems from the fact that the Lorentz boosts have opposite signs in the two conjugated representations, as shown in equations (5) and (6) and in Figure 1.

In Figure 2 the PSPs of a fiber with no PDL have been represented in solid black. They are orthogonal because the imaginary part of the Lorentz Group generator, see equation (4), is zero as explained in Section 2. When there is PDL in the optical fiber, the PSPs become complex and are no longer orthogonal to each other. They have been represented in solid blue in Figure 2. The input spinor $\xi_T$ is transformed as it propagates along the fiber to become the output spinor $\xi_R$. This is schematically represented in Figure 2. The dashed line joining the tips of the PMD vectors (in solid black) and the tips of the complex PSPs (in solid blue) and represent the imaginary parts of the PSPs. The arrows in solid red represent the PSPs in the conjugate representation. Notice that the conjugated spinors can be orthogonal at both ends of the fiber, and therefore related by a rotation, because the PSPs are different in both representations.

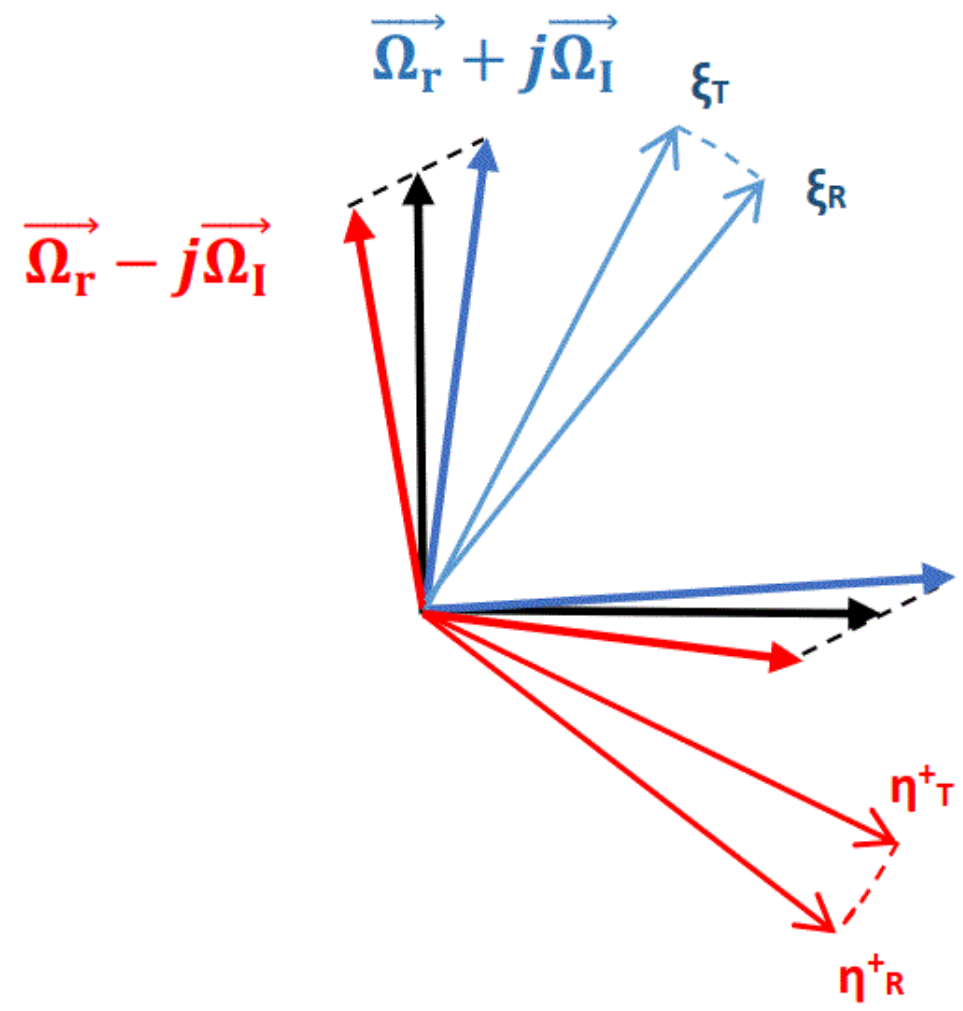

**Figure 2**: A single rotation joins the two input spinors with the two output spinors.

Two mutually orthogonal SOPs (the transmitted SOP and its conjugated SOP) at the fiber input can only be related to two mutually orthogonal SOPs at the fiber output (the received SOP and its conjugated SOP) by a rotation which proves that the PDL part of the impairments cancels when the correct mathematical framework is used to describe the problem.

An example of how these mathematical results could be used to enhance the estimation process of the transmitted states of polarization would be:

> At the receiver end of communication link, both "transmitted" SOPs could be estimated using the received and the conjugated SOPs and these estimation processes should run in parallel. The estimation processes should be refined by imposing the orthogonality between the two estimates of the two "transmitted" SOPs.

Notice that it has been assumed that the conjugated SOP can be exactly obtained by an ideal phase conjugator. No attempt has been made to describe the consequences of a less than perfect reconstruction of the conjugated SOP because of the theoretical nature of this paper.

## 5. PHYSICAL INTERPRETATION

The previous sections assume that polarization issues in optical communications are described by a mathematically consistent theory. Mathematical consistency prescribes the existence of two kinds of polarization states, the regular representation ξ, and the conjugate representation $\eta^{\dagger}$, that are mutually related by optical phase conjugation. Moreover, it also prescribes that these two spinors transform differently in an optical fiber. The regular spinor transforms as equation (5), while the conjugate spinor transforms as equation (6). These two equations ensure that polarization transformations in optical communications are complex analytic functions. However, this mathematical prescription has physical consequences that seem strange. An optical fiber seems to be described by two different SL(2,C) matrices depending on which representation the SOP belongs to. In this section, this dual description is explained in physical terms.

If the polarization dependent losses of the fiber are negligible, equations (5) and (6) coincide and the optical fiber transforms spinors belonging to the regular and conjugate representations in the same manner. Polarization mode dispersion alone does not distinguish between spinors belonging to either the regular or conjugate representation. The mathematical formulation described in this paper, although useful, is not indispensable.

In Jones calculus, the polarization of light is defined by **the complex envelope wave** of the electrical field:

$$\xi = \begin{pmatrix} \mathbf{E_x(t)} \\ \mathbf{E_y(t)} \end{pmatrix} = \begin{pmatrix} \mathbf{E_{0x}(t)e^{j\phi_x(t)}} \\ \mathbf{E_{0y}(t)e^{j\phi_y(t)}} \end{pmatrix}$$

In optical communications both the amplitude and the phase of the complex envelope depend on the symbol that is being transmitted so, they are both function of time (although their characteristic time of change is slow compared to the period of the optical carrier). Both the amplitude and the phase terms are used to encode information. Phase terms, $\phi_x$ and $\phi_y$ represent independent deviations from the phase reference of the electromagnetic wave (that must be retrieved from the optical carrier at the receiving end by means of an optical phase locked loop). The number of independent degrees of freedom is four, two amplitudes and two phases or two real and two imaginary parts. If we describe SOPs in Stokes formalism the number of degrees of freedom is just three (the three components of a vector in a 3D real space). Stokes formalism cannot describe SOPs in optical communications simply because it cannot accommodate four degrees of freedom.

If we conjugate the complex envelope of this electric field, the encoded information carried by the wave changes to:

$$\mathbf{E} \rightarrow \begin{pmatrix} \mathbf{E_{0x}(t)e^{-j\phi_x(t)}} \\ \mathbf{E_{0y}(t)e^{-j\phi_y(t)}} \end{pmatrix}$$

Notice that the real part of the complex envelope does not change in this transformation and that the imaginary part changes sign. To describe this transformation, we must be able to change independently the phase terms of both components of the electrical field. The device that describes this transformation is an ideal vector phase conjugating mirror [15]. The transformation that we call optical phase conjugation (it is a physical transformation) implements a different, although closely related, transformation:

$$\xi \rightarrow \boldsymbol{\eta}^{\dagger} = \begin{pmatrix} \mathbf{E_{0y}(t)e^{-j\phi_y(t)}} \\ -\mathbf{E_{0x}(t)e^{-j\phi_x(t)}} \end{pmatrix}$$

The optical setup that carries out this transformation is a π/4 Faraday rotator followed by an ideal vector phase conjugating mirror. Please note that the conjugation part would remove, in back-propagation, the PMD part of the polarization impairments and the Faraday rotator the PDL part (if the optical waves could back-propagate). If this transformation is obtained at the transmitting end, it represents a parity inversion operation. If it is calculated at the receiving end, it represents a time inversion operation. Since in this model these two discrete symmetries are identical, we conclude that the SOPs belonging to the conjugate representation are related, not by the SL(2,C) matrix (5), but by the matrix (6). These two representations clearly break the parity symmetry of Electromagnetism. How is this possible? The answer is that although Electromagnetism conserves parity, optical communication systems explicitly break this symmetry. Optical isolators and circulators make it impossible for optical waves to propagate in the opposite sense (from the receiving to the transmitting end) so that the time inversion symmetry is broken because the polarization distortions introduced by the fiber cannot be inverted. Since in this model both parity and time inversion are the same symmetry, parity inversion symmetry is also explicitly broken.

The mathematical model presented in this paper seems to suggest that PDL is caused by the non-reciprocal components (optical isolators and circulators) present in the optical communication system. In these components there is a large longitudinal magnetic field that seems to interact with the residual angular momentum of the propagating waves. Optical phase conjugation inverts this residual angular momentum and, consequently, the maximum and minimum attenuation axes of these components seem to exchange their roles.

When the preceding paragraph was written, this assertion was pure speculation because there was no reported physical effect that corroborated it. However, a recent publication [16] by two researchers from the Hebrew University of Jerusalem, B. Assouline and A. Capua, proves that the interaction between the angular momentum of the light and the large longitudinal magnetic field present in Faraday rotators accounts for 70% of the rotation at infrared wavelengths. Since the Landau–Lifshitz–Gilbert (LLG) equation used in their report includes a dissipative coefficient, the α Gilbert parameter, there seems to be a physical explanation that corroborates this speculation. Please, remember that optical isolators and circulators are non-reciprocal devices because they include Faraday rotators. Note that a long-haul optical communication system typically contains many isolators and circulators.

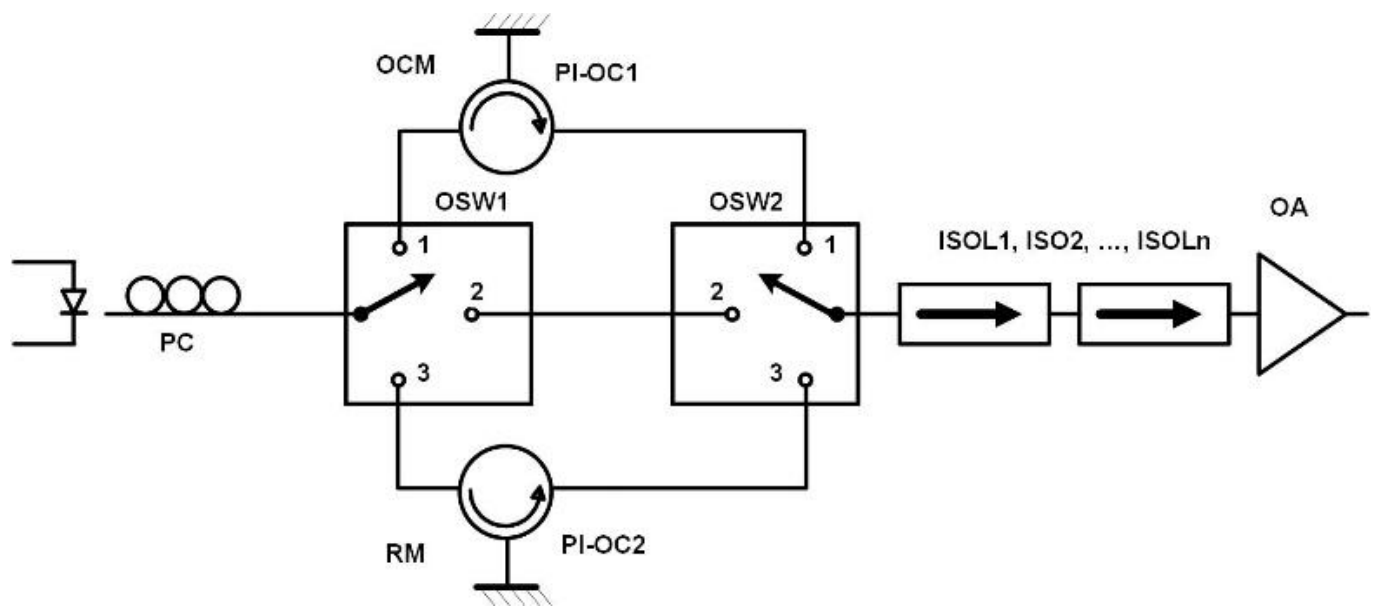

**Figure 3:** Proposed experimental setup.

## 6. EXPERIMENTAL SETUP

The mathematical model proposed in this paper states that two different matrices are needed to describe the polarization behavior of a fiber optic. Since this idea looks dubious, I propose an experimental set up (Figure 3) intended to either validate or falsify the proposed mathematical model.

In this setup, the SOP of light generated by a highly coherent semiconductor laser is arbitrary changed by a polarization controller PC and connected to a one input-three output optical switch $OSW_1$. The three outputs of this optical switch are directed to either a conjugate mirror (1), or a regular mirror (3) or an optical fiber whose beat length is very long (2). The outputs of the conjugate mirror (OCM), regular mirror (RM) and the long beat-length fiber are connected to the three inputs of a similar optical switch, $OSW_2$. The output light wave propagates along a set of "n" optical isolators $ISOl_1$, $ISOL_2$, …, $ISOL_n$ (in Figure 3 only two isolators have been represented). The polarization independent optical losses are compensated in an optical amplifier OA. The optical circulators depicted in Figure 3 should not change the SOPs of the light so they should be polarization-insensitive optical circulators (PI-$OC_1$ and PI-$OC_2$).

The mathematical model proposed in this paper predicts that SOP of the output waves when the optical switches are in positions 1 and 2 should be mutually orthogonal independently of the SOP at the output of the polarization controller PC. This can only happen if the fiber polarization properties are described by two different SL(2,C) matrices, (5) and (6). However, no such systematic orthogonality should be observed when the optical switches are in positions 2 and 3. If this prediction is experimentally verified it would mean that the proposed mathematical model can be correct. If it is not, it would mean that it is not correct and the mathematical model used in optical communications to describe polarization issues would not be mathematically consistent. Please note that the proposed model directly stems from the assumption that the mathematics describing polarization in optical communications are consistent.

## 7. CONCLUSIONS

The idea of interpreting polarization changes as restricted Lorentz Transformations was well known in Optics but had been completely ignored in the field of optical communications. However, this idea provides the key to a correct mathematical description of polarization issues in this field. The polarization transformations that take place in optical communications can be regarded as restricted Lorentz transformations. The restricted Lorentz Group can be extended to the complete Lorentz by including the three discrete symmetries when optical phase conjugation is identified as the optical process that allows to have access to the conjugate representation of polarization. These two representations provide all the relevant information about combined PMD-PDL impairments. When both spinors are considered, PDL impairments can be minimized. This mathematical framework proves to be essential to understand the polarization transformations that take place in fiber optic communication systems and to realize what can be and cannot be done in these systems. The PDL can be minimized even though each input spinor is related to each output spinor by (5) and (6) that are not rotational relationships. The joint motion of both input spinors is just rotation.

**Acknowledgments.** I would like to express my sincere gratitude to Prof. Vicente Baena of the University of Seville and Prof. Michael Connelly of the University of Limerick for the many discussions held in the last twenty years in seemingly disparate topics of Communication Engineering, Physics and Mathematics.

## References

1 , L. Leng, S. Stulz, B. Zhu, L.E. Nelson, B. Edvold, L. Gruner-Nielsen, S. Radic, J. Centanni, and A. Gnauck, "1.6-Tb/s (160x10.7 Gb/s) transmission over 4000 km of nonzero dispersion fiber at 25-GHz channel spacing," IEEE Photon. Technol. Lett. 15, 1153-1155 (2003).

2 H. Bülow and S. Lanne, "PMD compensation techniques," In: Galtarossa, A., Menyuk, C.R. (eds) Polarization Mode Dispersion. Optical and Fiber Communications Reports, vol 1. Springer, New York, NY, (2004).

3 B. Huttner, C. Geiser and N. Gisin, "Polarization-induced distortions in optical fiber networks with polarization-mode dispersion and polarization dependent losses," IEEE Journal of Selected Topics in Quantum Electronics, vol. 6, no. 2, pp. 317-329 (2000).

4 T. Tudor, "Lorentz Transformations, Poincaré Vectors and Poincaré Sphere in Various Branches of Physics," Symmetry 5, 233-252 (2013).

5 A. Z. Goldberg, P. de la Hoz, G. Björk, A. B. Klimov, M. Grassl, G. Leuchs and L. L. Sánchez-Soto, "Quantum concepts in optical Polarization," Advances in Optics and Photonics vol. 13, Issue 1, 1-73 (2021).

6 Y. S. Kim, "Lorentz group in polarization optics," Journal of Optics B: Quantum and Semiclassical Optics, Vol. 2, no 2, 1-5 (2000).

7 J. P. Gordon Lamont and H. Kogelnik, "PMD fundamentals: Polarization mode dispersion in optical fibers," PNAS vol. 97, no 9, 4541-4550 (2000).

8 J. N. Damask, "Polarization Optics in Telecommunications," Springer, (2005).

9 A. Zee, "Group Theory in a Nutshell for Physicists", Princeton University Press, (2016).

10 H. Georgi, "Lie Algebras in Particle Physics", Westview Press, (1999).

11 L. H. Ryder, "Quantum Field Theory", 2nd Edition, Cambridge University Press, (1985).

12 G. Costa and G. Fogli, "Symmetries and Group Theory in Particle Physics", Springer, (2012).

13 R. A. Fisher, "Optical Phase Conjugation", Academic Press, (1983).

14 C. L. Janer and M. J. Connelly, "Optical phase conjugation technique using four-wave mixing in semiconductor optical amplifier", El. Lett., vol. 47, Issue 12, (2011).

15 Mao Ye, Peng-Ye Wang, Jian-Hua Dai and Hong-Jun Zhang, "Vector phase conjugation by degenerate four-wave mixing in a nematic liquid crystal film", Optics Communications, Volume 104, Issues 1–3, (1993).

16 B. Assouline and A. Capua, "Faraday effects emerging from the optical magnetic field", *Sci Rep* **15**, 39566 (2025).